\documentclass[aps,pre,amsmath,amssymb,showpacs,preprint]{revtex4-1}
\usepackage{bm,graphicx}
\newcommand{\ud}{\mathrm{d}}

\newcommand{\ui}{\mathrm{i}}

\begin{document}

 \title{Anomalous~diffusion for~inertial~particles under~gravity in~parallel~flows}
 
 \author{Marco \surname{Martins Afonso}}
 \affiliation{Laboratoire~de~M\'ecanique,~Mod\'elisation~et~Proc\'ed\'es~Propres, CNRS~UMR~7340, Aix-Marseille~Universit\'e, Ecole~Centrale~Marseille, 38~rue~Fr\'ed\'eric~Joliot-Curie, 13451~Marseille~cedex~13, France}
 \email{marcomar@fisica.unige.it}

 \date{\today}

 \begin{abstract}
  We investigate the bounds between normal or anomalous effective diffusion
  for inertial particles transported by parallel flows.
  The infrared behavior of the fluid kinetic-energy spectrum,
  i.e.\ the possible presence of long-range spatio-temporal correlations,
  is modeled as a power law by means of two parameters,
  and the problem is studied as a function of these latter.
  Our results, obtained in the limit of weak relative inertia,
  extend well-known results for tracers and
  apply to particles of any mass density, subject to gravity and Brownian diffusion.
  We consider both steady and time-dependent flows, and cases of both vanishing and finite particle sedimentation.
 \end{abstract}
 

 \maketitle

 \section{Introduction}
 
 The concept of particle diffusion is ubiquitous in dynamical systems and fluid mechanics \cite{CFPV91,AM91,AV95}.
 Most classical works \cite{MK99} dealt with Lagrangian tracer particles, but some studies \cite{R88,CFPV90,PS05,MAMMG12}
 have also focused on particles endowed with inertia (relative to the fluid), such as drops in gases, bubbles in liquids,
 and more generally aerosols in fluids. In \cite{FJBE06} this problem was attacked in the phase space.
 
 If a statistical description is introduced, whenever the central-limit theorem holds a normal diffusion process takes place,
 i.e.\ the mean square displacement of particles follows
 \begin{equation} \label{gam}
  \langle|\bm{r}(t)-\bm{r}(0)|^2\rangle\sim t^{\gamma}
 \end{equation}
 with $\gamma=1$, at long times (and thus large scales). The proportionality coefficient is named \emph{eddy diffusivity},
 and its value can be by many orders of magnitude different (typically, larger) than its Brownian or molecular counterpart.
 It is however possible to find exceptions to this normal picture,
 in which case $\gamma\neq1$ \cite{BCVV95,CCMVV98,CMMV99,ACMV00} and the term \emph{anomalous diffusion} is used.
 Namely, one can have subdiffusion --- due to trapping processes --- if $\gamma<1$, or superdiffusion if $\gamma>1$.
 In this latter situation, one finds a divergence in the eddy-diffusivity coefficient,
 or in its correspondent effective-diffusivity tensor if the full tensorial problem
 is investigated in the Eulerian framework for the physical-space concentration.
 
 The aim of the present work is to identify bounds separating situations of normal or anomalous diffusion,
 for inertial-gravitational-Brownian particles advected by a \emph{parallel flow}, which points always and everywhere in the same direction.
 This task is performed in terms of the behavior of the fluid velocity spectrum in the infrared region, i.e.\ very small wave numbers or frequencies
 in the Fourier space, which describe the possible presence of long-range spatio-temporal correlations
 that in turn may help in breaking the process of normal diffusion.
 Our investigation takes into account particle inertia, gravity, Brownian diffusivity and added mass
 (it is thus valid for any ratio of the mass densities), and can be divided in different chapters
 according to three main discriminating factors: first, tracers or weakly-inertial particles;
 second, vanishing or finite sedimentation in the limit of small inertia; third, steady or time-dependent flows.
 
 The paper is organized as follows.
 In section~\ref{equa} we recall the relevant equations
 for the problem under consideration from the existing scientific literature.
 Section~\ref{andi} is devoted to present the bounds on the parameters for the presence or absence of anomalous diffusion.
 Conclusions and perspectives follow in section~\ref{cope}.
\textsc{The appendix~\ref{appe} shows the regularization procedure and the relation between the scaling exponents.}

 \section{Equations} \label{equa}

 Let us consider a dilute suspension of identical, spherical inertial particles
 (of radius $R$ and mass density $\rho_{\mathrm{p}}$),
 subject to the gravity acceleration $\bm{g}$ and to the Brownian diffusivity $\kappa$,
 advected by an incompressible $d$-dimensional ($d=2,3$) zero-mean fluid flow
 of mass density $\rho_{\mathrm{f}}$ and kinematic viscosity $\nu$.
 Under some simplifying approximations (discussed e.g.\ in \cite{MAMMG12} and references therein),
 their position $\bm{\mathcal{X}}(t)$ and covelocity
 $\bm{\mathcal{V}}(t)\equiv\dot{\bm{\mathcal{X}}}(t)-\beta\bm{u}(\bm{\mathcal{X}}(t),t)$
 evolve according to \cite{MR83,G83}:
 \begin{equation} \label{dyn}
  \left\{\begin{array}{l}
   \dot{\bm{\mathcal{X}}}(t)=\bm{\mathcal{V}}(t)+\beta\bm{u}(\bm{\mathcal{X}}(t),t)\\
   \dot{\bm{\mathcal{V}}}(t)=\displaystyle-\frac{\bm{\mathcal{V}}(t)-(1-\beta)\bm{u}[\bm{\mathcal{X}}(t),t]}{\tau}+(1-\beta)\bm{g}+\frac{\sqrt{2\,\kappa}}{\tau}\bm{\eta}(t)\;,
  \end{array}\right.
 \end{equation}
 where $\bm{\eta}(t)$ is the standard white noise and $\tau=R^2/3\nu\beta$ is the Stokes response time.
 Here, we have defined the adimensional coefficient $\beta\equiv3\rho_{\mathrm{f}}/(\rho_{\mathrm{f}}+2\rho_{\mathrm{p}})$
 based on the density ratio, ranging from $\beta=0$ for very heavy particles to $\beta=3$ for very light ones
 (with $\beta=1$ for neutrally-buoyant particles, such as tracers).
 We can also introduce the Stokes number $\mathrm{St}\equiv\tau/(\ell/U)$
 --- $\ell$ and $U$ being the characteristic length and speed scales of the carrier flow ---
 which measures the importance of the relative inertia between particles and fluid.
 Notice that the real particle velocity $\dot{\bm{\mathcal{X}}}(t)=\bm{\mathcal{V}}(t)+\beta\bm{u}(\bm{\mathcal{X}}(t),t)$
 equals the covelocity only for very heavy particles, while in the other cases the discrepancy
 simply represents an easy way to take the added-mass effect into account.
\textsc{The other two relevant nondimensional numbers of the problem are due to P\'eclet,
 $\mathrm{Pe}\equiv\ell U/\kappa$, and to Froude, $\mathrm{Fr}\equiv U/\sqrt{g\ell}$.}
 In still fluids, the particle bare settling velocity is $\bm{w}_*=(1-\beta)\bm{g}\tau$
\textsc{($=(1-\beta)\mathrm{St}\mathrm{Fr}^{-2}$, in units of $U$, along the vertical).}

 Straightforward from (\ref{dyn}), the generalized Fokker--Planck (or Kramers, or forward Kolmogorov) equation
 for the phase-space particle density $\rho(\bm{x},\bm{v},t)$ reads
 \cite{C43,G85,R89,V07}:
 \begin{equation} \label{fp}
  \left\{\frac{\partial}{\partial t}+\frac{\partial}{\partial x_{\mu}}\left(v_{\mu}+\beta u_{\mu}\right)+\frac{\partial}{\partial v_{\mu}}\left[\frac{(1-\beta)u_{\mu}-v_{\mu}}{\tau}+(1-\beta)g_{\mu}\right]-\frac{\kappa}{\tau^2}\frac{\partial^2}{\partial v_{\mu}\partial v_{\mu}}\right\}\rho=0\;.
 \end{equation}
 Let us denote with $\mathcal{L}$ the differential operator specified by the curly brackets in (\ref{fp}),
 so that the equation becomes $\mathcal{L}\rho(\bm{x},\bm{v},t)=0$
 endowed with the appropriate initial condition $\rho(\bm{x},\bm{v},0)$.
 In the presence of a flow, the particle terminal velocity is simply the particle velocity averaged on the phase space with this weighting density,
 \begin{equation} \label{tv}
  \bm{w}\equiv\!\int_0^{\mathcal{T}}\!\frac{\ud t}{\mathcal{T}}\int\!\ud\bm{x}\int\!\ud\bm{v}\,[\bm{v}+\beta\bm{u}(\bm{x},t)]\rho(\bm{x},\bm{v},t)\;.
 \end{equation}

 We are now interested in analyzing the problem at a spatial scale $L\gg\ell$ and at a very long temporal scale, say $\gg\mathcal{T}$,
 where $\mathcal{T}$ can be thought of as the typical advective time scale $\ell/U$.
 By means of a multiple-scale expansion \cite{BLP78,BO78,PS07} in the scale-separation parameter, $\epsilon\equiv\ell/L\ll1$,
 it was shown in \cite{PS05,MAMMG12} that, in the frame of reference moving with the terminal velocity $\bm{w}$,
 the large-scale, long-time behavior of the particle concentration can be described by means of a diffusion equation.
 Namely, introducing the slow variables $\bm{X}\equiv\epsilon\bm{x}$ and $T\equiv\epsilon^2t$
 (which are to be considered as independent from the corresponding fast variables $\bm{x}$ and $t$,
 and become $O(1)$ only for very large values of these latter), and factorizing the particle concentration
 as $\rho(\bm{x},\bm{X},\bm{v},t,T)=p(\bm{x},\bm{v},t)P(\bm{X},T)$, it was proved that the slow component $P(\bm{X},T)$ satisfies:
 \begin{equation} \label{de}
  \frac{\partial}{\partial T}P(\bm{X},T)=K_{\lambda\mu}\frac{\partial^2}{\partial X_{\lambda}\partial X_{\mu}}P(\bm{X},T)\;.
 \end{equation}
 The effective-diffusivity tensor $\mathsf{K}$ in (\ref{de}) is given by
 \begin{equation} \label{ed}
  K_{\lambda\mu}=-\!\int_0^{\mathcal{T}}\!\frac{\ud t}{\mathcal{T}}\int\!\ud\bm{x}\int\!\ud\bm{v}\,\left[v_{\mu}+\beta u_{\mu}(\bm{x},t)-w_{\mu}\right]\sigma_{\lambda}(\bm{x},\bm{v},t)+\textrm{symm.}(\lambda\leftrightarrow\mu)\;,
 \end{equation}
 with the auxiliary vector $\bm{\sigma}$ in (\ref{ed}) satisfying the cell problem
 \begin{equation} \label{cp}
  \mathcal{L}\bm{\sigma}(\bm{x},\bm{v},t)=-\left[\bm{v}+\beta\bm{u}(\bm{x},t)-\bm{w}\right]p(\bm{x},\bm{v},t)
 \end{equation}
 endowed with vanishing initial condition $\bm{\sigma}(\bm{x},\bm{v},0)=\bm{0}$.\\
 A complete solution of the problem would thus require to solve (\ref{fp}) for $\rho(\bm{x},\bm{v},t)$,
 to plug it into (\ref{tv}) in order to find the correct frame of reference in which to investigate diffusion
 as free from any ballistic degree of freedom \cite{PS05}, then to solve (\ref{cp}) for $\bm{\sigma}(\bm{x},\bm{v},t)$,
 and finally to plug this latter into (\ref{ed}).
 
 In \cite{MAMMG12} an expansion at small particle inertia was also performed,
 i.e.\ when the inertial particles weakly deviate with respect to the
 underlying fluid trajectories ($\mathrm{St}\ll1$).
 The diffusivity tensor was thus expressed through an expansion in the Stokes number:
 \begin{equation} \label{exp}
  K_{\lambda\mu}=K^{(0)}_{\lambda\mu}+\mathrm{St}K^{(1)}_{\lambda\mu}+O(\mathrm{St}^2)\;.
 \end{equation}
\textsc{A perturbative expansion in $\mathrm{Pe}$ was made in \citep{BCVV95} in a different asymptotics.}

 Now, two situations may emerge, according to the constitutive relationship assigned between sedimentation and inertia.
 Let us first consider the case where the nondimensionalized bare settling velocity, $w_*/U$, vanishes for vanishing particle inertia.
 For parallel flows $u_{\mu}(\bm{x},t)=\delta_{\mu1}u(x_2,\ldots,x_d,t)$, it was found in \cite{BCVV95} (generalizing a result from \cite{Z82})
 that the tracer limit $\mathsf{K}^{(0)}$ of the eddy diffusivity can be expressed as:
 \begin{equation} \label{trac}
  K^{(0)}_{\lambda\mu}=\kappa\left[\delta_{\lambda\mu}+\delta_{\lambda1}\delta_{\mu1}\!\int\!\ud\vec{q}\int\!\ud\omega\,\mathcal{U}(\vec{q},\omega)\frac{q^2}{\omega^2+\kappa^2q^4}\right]\;.
 \end{equation}
 Here, $\mathcal{U}(\vec{q},\omega)$
 is the energy spectrum density --- with dimensions (length/time)$^2\times$(length$^{d-1}$\linebreak$\times$time) = length$^{d+1}$/time ---
 obtained via a Fourier transform in both time and the $d-1$
 spatial coordinates on which the flow depends
 (so that $\omega$ is the angular frequency, and $\vec{q}$ is the $(d-1)$-dimensional wave number vector,
 which obviously reduces to a scalar in the two-dimensional case).
\textsc{Note that we adopt a frame of reference with the $x_1$ axis pointing along the flow,
 therefore gravity is in general not aligned with any of the --- two or three --- Cartesian directions.
 As we focus on isotropic cases, the energy spectrum density is alternatively defined as
 \begin{equation} \label{spe}
  \mathcal{U}'(q,\omega)\propto|\vec{q}|^{d-2}\mathcal{U}(\vec{q},\omega)\;,
 \end{equation}
 in order to take the Jacobian factor into account, and such that the kinetic energy per unit mass is equivalently
 $\int\!\ud\vec{q}\int\!\ud\omega\,\mathcal{U}(\vec{q},\omega)=\int\!\ud q\int\!\ud\omega\,\mathcal{U}'(q,\omega)$.}\\
 Reference \cite{MAMMG12} further showed that for this class of flows the leading correction $\mathsf{K}^{(1)}$ at small inertia is:
 \begin{equation} \label{corr}
  K^{(1)}_{\lambda\mu}=\delta_{\lambda1}\delta_{\mu1}\frac{\ell}{U}\!\int\!\ud\vec{q}\int\!\ud\omega\,\mathcal{U}(\vec{q},\omega)\frac{(1-\beta)\omega^2+(3-\beta)\kappa^2q^4}{2(\omega^2+\kappa^2q^4)}\;.
 \end{equation}
 Notice that (\ref{corr}) is an inertial additive correction to (\ref{trac}) under a perturbative scheme.
 If (\ref{trac}) converges and tracers diffuse normally, a convergent (\ref{corr}) suggests that
 normal diffusion also holds for inertial particles, while a divergent (\ref{corr}) may indicate
 an inertia-driven anomaly.  (We say \emph{may indicate}, and not ``does indicate'', because a 
 regularizing normalization cannot be excluded for the full resummation (\ref{exp}).)
 If (\ref{trac}) diverges and tracers diffuse anomalously, a convergent (\ref{corr}) means that
 inertia does not change this picture at its leading order, while
 no conclusion can be drawn a priori for
 a divergent (\ref{corr}) except in some specific cases.

 An alternative point of view consists in considering sedimentation as a finite effect even for vanishing inertia,
 which is the case when the nondimensional bare settling velocity keeps finite for $\mathrm{St}\to0$ \cite{M87}.
 In this case, for parallel flows, gravity plays the role of a constant drift
 already at the zeroth-order in the Stokes number \cite{MAMMG12}:
 \begin{equation} \label{max}
  K^{(0)}_{\lambda\mu}=\kappa\left\{\delta_{\lambda\mu}+\delta_{\lambda1}\delta_{\mu1}\!\int\!\ud\vec{q}\int\!\ud\omega\,\mathcal{U}(\vec{q},\omega)\frac{q^2}{[\omega+(1-\beta)\tau\bm{g}\cdot\bm{q}]^2+\kappa^2q^4}\right\}\;,
 \end{equation}
 where $\bm{q}=(0,\vec{q})$ is a usual $d$-dimensional vector.
 This sweeping effect makes the role of the inertial correction $\mathsf{K}^{(1)}$ by far less interesting in this case,
 therefore we will not investigate it. It is worth underlining that, while the activation of inertia --- for vanishing sedimentation ---
 implies an additive correction (\ref{corr}) to (\ref{trac}) (which thus remains important), here the activation of a finite sedimentation
 means that (\ref{max}) completely replaces (\ref{trac}), so that the latter is not relevant any longer.
 This case can be compared to the one investigated in \cite{M97} about tracer diffusion in the presence of a large-scale sweeping flow.
 The role of parallel streaming flows was also studied in \cite{MMV05}.
\textsc{A finite sedimentation thus corresponds to a constant large-scale flow, except for the fact that this latter
 is clearly not originated from inertia. Its interplay with inertia, i.e.\ the study of the leading impact of inertia on eddy diffusivity,
 could e.g.\ stem from the analysis of the aforementioned correction $\mathsf{K}^{(1)}$, which is not done here for the sake of simplicity.
 On the other hand, a different picture could arise if a space-time-dependent large-scale flow is added to the right-hand side of (\ref{dyn}),
 which might be interesting for applications in the realm of large-eddy simulations. If this new component is concentrated on spatial
 and temporal scales $S_{\mathrm{spat}}$ and $S_{\mathrm{temp}}$, respectively, a preasymptotic regime is then met when studying the problem
 at scales $\gg(\ell,\mathcal{T})$ and $\ll(S_{\mathrm{spat}},S_{\mathrm{temp}})$; a full homogenization into a purely-diffusive (normal or
 anomalous) problem --- with a consequent new renormalization of the eddy diffusivity into a finite or infinite value ---
 is generally possible when considering scales sufficiently larger than $(S_{\mathrm{spat}},S_{\mathrm{temp}})$ themselves.}

 \section{Anomalous diffusion} \label{andi}

 Let us now consider the behavior of the particle effective diffusivity as a function of the fluid velocity spectrum.
 The crucial point is how the latter behaves near the origin, i.e.\ the possible presence of long-range correlations
 in both the spatial and the temporal domains. Notice that we are not interested in the behavior at large wave number
 nor at high frequency, where ultraviolet cut-offs will take place,
\textsc{but only in the infrared one.}

\textsc{It is worth noticing that the analysis of possible anomalies here relies on the presence
 of a spatial velocity spectrum unbounded in the infrared region, i.e.\ with wave lengths extending to infinity.
 This poses two problems, namely the definite non-periodicity of the fluid flow under investigation,
 and the impossibility of defining an observation length much larger than any spatial scale possessed by the velocity field.
 The first difficulty can be overcome by recalling that the results of \cite{BCVV95,MAMMG12}
 also apply for non-periodic but random (stationary and homogeneous) velocity fields,
 by appropriately reformulating the space-time integrals as statistical averages \cite{AM91}.
 The second point requires the use of a regularization procedure, as explained in \cite{BCVV95}:
 this is investigated in the appendix \ref{appe}.}

 \subsection{Steady flows} \label{stfl}
 
 Let us start with the case of steady flows, for which the temporal part
 of the spectrum is simply a centered Dirac delta. Let us then assume a power-law form for the spatial part (isotropic in the relevant $d-1$
 dimensions) with scaling exponent $\alpha$, for $|\vec{q}|$ small enough --- say $|\vec{q}|<Q$ for a suitable $Q$:
 \begin{equation} \label{alfa}
  \mathcal{U}(\vec{q},\omega)\sim|\vec{q}|^{\alpha}\delta(\omega)\;.
 \end{equation}
\textsc{If the modified spectrum from (\ref{spe}) is used, then $\mathcal{U}'(q,\omega)\sim|\vec{q}|^{\alpha'}\delta(\omega)$ with
 \begin{equation} \label{alf}
  \alpha'=\alpha+d-2\;.
 \end{equation}
 See \cite{C07} for an interesting analysis of the role played by velocity fields with power-law spectra in modeling turbulent flows.}

 Our task now consists in replacing (\ref{alfa}) into the expressions of the eddy diffusivity tensor,
 namely in the $K_{11}$ component, and to study the behavior of the corresponding integrals for small values of $q$.
 The smaller the exponent $\alpha$, the heavier the relevance of long-range spatial correlations,
 i.e.\ the higher the probability of anomalous diffusion. The temporal integrals are of course trivial
 because of the Dirac delta, therefore one can just study the spatial form of the integrand with $\omega=0$.
 
 \subsubsection{Case of vanishing terminal velocity} \label{cvtv}
 
 In the case of vanishing terminal velocity, (\ref{trac}) reproduces a well-known result \cite{BCVV95}:
 in cylindrical/spherical coordinates, the integral (whose Jacobian is $\propto q^{d-2}$
 and whose eventual angular part is trivial) takes the form
 $$\kappa\!\int_0^Q\ud q\,q^{d-2}q^{\alpha}\frac{q^2}{\kappa^2q^4}\propto\!\int_0^Q\ud q\,q^{\alpha+d-4}\;,$$ 
 which exists for
 \begin{equation} \label{1}
  \alpha>3-d;.
 \end{equation}
 Smaller values of $\alpha$ denote anomalous diffusion, namely a super-diffusive behavior of the particles.
 Note that, here and in what follows, the inequalities are strict,
 in the sense that if the parameter exactly equals the bound then logarithmic divergences (anomalies) occur.
 All these bounds can
\textsc{also be found more elegantly and rigorously}
 by means of the Mellin transform,
 but here we prefer to present them in a more physical and
\textsc{easier-to-interpret}
 fashion.

 The additive (see (\ref{exp})) leading correction due to inertia, i.e.\ the integral
 $$\frac{\ell}{U}\!\int_0^Q\ud q\,q^{d-2}q^{\alpha}\frac{(3-\beta)\kappa^2q^4}{2\kappa^2q^4}\propto\!\int_0^Q\ud q\,q^{\alpha+d-2}$$
 from (\ref{corr}), gives a threshold
 \begin{equation} \label{0}
  \alpha>1-d\;,
 \end{equation}
 which is less restrictive than (\ref{1}), because the right-hand side $1-d$ is always smaller than $3-d$.
 In other words, for values of $\alpha$ larger than $1-d$ this is a finite correction to a finite or infinite leading order,
 which makes no change in terms of normality or anomaly.
 On the contrary, for smaller $\alpha$ this is an infinite correction to an infinite leading order,
 therefore no ultimate conclusion could be drawn a priori because one might not exclude the possibility of having a renormalization
 upon taking into account all the terms in expansion (\ref{exp}) (as in asymptotic series);
 however, if we remind that anomaly is more and more likely for smaller and smaller exponents $\alpha$,
 we can deduce that the activation of inertia does not modify the anomalous character of diffusion in this range.

 \subsubsection{Case of finite terminal velocity} \label{cftv}

 For finite terminal velocity, on the other hand, one has to study (\ref{max}).
 (Let us exclude the cases where the flow is aligned vertically ($\bm{g}\perp\bm{q}\ \forall\vec{q}$)
 or the particles are neutrally buoyant ($\beta=1$),
 which would exactly give back (\ref{trac}) and thus (\ref{1}).)
 The term at denominator containing gravity can radically change the power balance in $q$ as long as it is nonzero,
 because for finite prefactors $q^2\gg q^4$ at small $q$, and thus the term containing Brownian diffusivity
 may simply be seen as a regularization \cite{BCVV95} acting only to avoid zeros at denominator.
 It is therefore crucial to investigate in detail the geometric aspect of the problem,
 in particular if and how the integration domain spans $\bm{q}$'s perpendicular to $\bm{g}$.\\
 In $d=2$ this orthogonality never occurs, because the integral is in fact one dimensional,
 and its rewriting in radial (i.e., absolute-value) form,
 $$\kappa\!\int_{-Q}^Q\ud q\,q^{d-2}|q|^{\alpha}\frac{q^2}{(1-\beta)^2\tau^2g^2\cos^2(\pi/2-\Theta)q^2+\kappa^2q^4}\sim\frac{2\kappa}{(1-\beta)^2\tau^2g^2\sin^2(\Theta)}\!\int_0^Q\ud q\,q^{\alpha}\;,$$
 simply results in twice an overall factor
 in terms of the angle $\Theta\neq0$ between the flow and the vertical direction.
 The Brownian-diffusivity regularization at denominator can safely be neglected,
 and the power balance for the consequent integration on $q$ shows an integrand
 proportional to $q^{\alpha+d-2}=q^{\alpha}$ at small $q$. The bound is thus $\alpha>-1$.\\
 On the contrary, for $d=3$, we introduce polar coordinates $(q,\theta)$ in the $\vec{q}$ integration plane,
 which is by definition perpendicular to the flow,
 and we denote with $\theta$ the integration angle computed starting by the projection of $\bm{g}$ onto this plane.
 The cosine of the angle between $\bm{g}$ and $\bm{q}$,
 which appears in the denominator of (\ref{max}), is given by the standard formula
 $\cos(\Theta)\cos(\vartheta)+\sin(\Theta)\sin(\vartheta)\cos(\theta)\mapsto\sin(\Theta)\cos(\theta)$,
 with variable $0\le\theta<2\pi$ and fixed $\Theta\neq0$ and $\vartheta=\pi/2$,
 this latter being the angle between $\bm{q}$ and the flow.
 The actual angular integration in $\theta$ must be performed
 keeping into account the regularizing term because the contribution from gravity vanishes at $\theta=\pi/2,3\pi/2$:
 \begin{eqnarray*}
  \kappa\!\int_0^Q\ud q\int_0^{2\pi}\ud\theta\,q^{d-2}q^{\alpha}\frac{q^2}{(1-\beta)^2\tau^2g^2\sin^2(\Theta)\cos^2(\theta)q^2+\kappa^2q^4}\\
  =2\pi\!\int_0^Q\ud q\,\frac{q^{\alpha}}{\sqrt{(1-\beta)^2\tau^2g^2\sin^2(\Theta)+\kappa^2q^2}}\;.
 \end{eqnarray*}
 The result is that, after the angular integration,
 the integrand for the radial integral behaves as $q^{\alpha+d-3}=q^{\alpha}$ for small $q$.
 Consequently, we obtain the following boundary for the presence of normal diffusion:
 \begin{equation} \label{2}
  \alpha>-1
 \end{equation}
 (i.e.\ the same critical value for both the two- and the three-dimensional case).\\
 Such a threshold is less restrictive than (\ref{1}), but it is important to point out that now inequality
 (\ref{2}) \emph{replaces} (\ref{1}) --- differently from the case of inertial particles with vanishing terminal velocity,
 for which (\ref{0}) is imposed on an additive term which sums up with the one ruled by (\ref{1}).
 The upshot is that some anomalously-diffusive cases for vanishing settling in (\ref{1}) (such as e.g.\ for $\alpha=-1/2$ for $d=2,3$)
 can be turned into normal-diffusion processes by the activation of a finite sedimentation.

 \subsection{Unsteady flows} \label{unfl}

 Let us now turn to the case of time-dependent flows, for which (for $|\vec{q}|<Q$ and $|\omega|$ small enough,
 say $|\omega|<\Omega$ for a suitable $\Omega$)
 we impose a power-law form also in the temporal part with scaling exponent $\zeta$:
 \[\mathcal{U}(\vec{q},\omega)\sim|\vec{q}|^{\alpha}|\omega|^{\zeta}\]
\textsc{(or equivalently $\mathcal{U}'(q,\omega)\sim|\vec{q}|^{\alpha'}|\omega|^{\zeta}$ along with (\ref{alf})).}

 The smaller the exponent $\zeta$, the heavier the relevance of long-time correlations,
 i.e.\ the higher the probability of anomalous diffusion.
 Now also the time integrals in the expressions of the eddy diffusivity must be performed with care.

 \subsubsection{Case of vanishing terminal velocity} \label{cvti}

\begin{figure}
 \centering
 \includegraphics{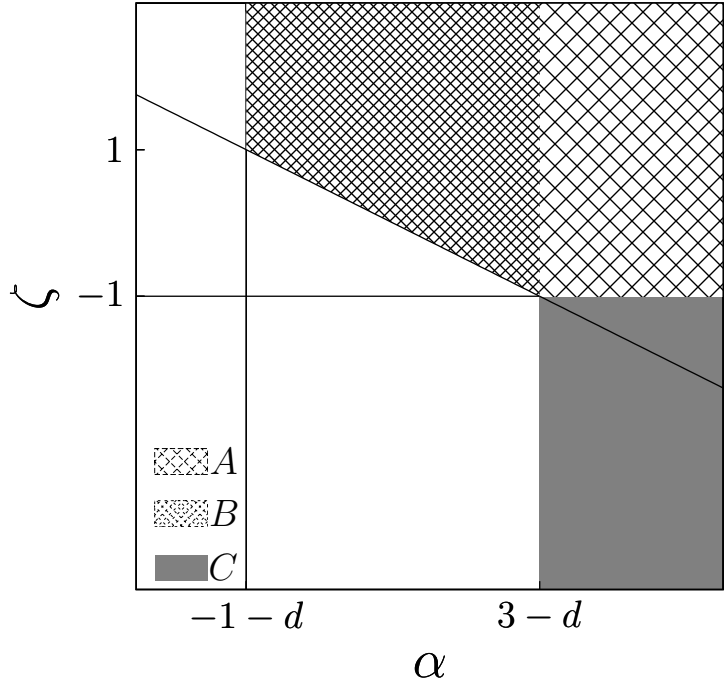}
 \caption{Sketch of diffusion anomaly for tracers, in the plane $\zeta$ vs.\ $\alpha$ (see (\ref{3})).
  Diffusion is normal only in meshed areas. The coarse-mesh area A has normal diffusion
  also for steady flows, while in the fine-mesh area B the normality is induced by the
  time dependence. The union of A and of the grey area C indicates the normal zone for time-independent flows (see (\ref{1})).}
 \label{fig1}
\end{figure}

 At the tracer level for vanishing sedimentation, (\ref{trac}), the double integral now takes the form
 $$\kappa\!\int_0^Q\ud q\int_{-\Omega}^{\Omega}\ud\omega\,q^{d-2}q^{\alpha}|\omega|^{\zeta}\frac{q^2}{\omega^2+\kappa^2q^4}\;,$$
 which is finite for
 \begin{equation} \label{3}
  \alpha>-1-d\quad\&\quad\zeta>-1\quad\&\quad\alpha+2\zeta>1-d\;.
 \end{equation}
 These three constraints define an open region in the upper right part of the plane $\zeta$ vs.\ $\alpha$,
 as shown in figure \ref{fig1}.
 In any case, comparing (\ref{3}) with (\ref{1}), the constraint on $\alpha$ is always less restrictive now:
 $\alpha>\max\{-1-d,1-d-2\zeta\}$ (where the right-hand side is always smaller than $3-d$
\textsc{if the constraint $\zeta>-1$ is satisfied).}
 This means that the introduction of a time dependence in the flow always causes anomalous diffusion for tracers
 in the presence of strong temporal coherence ($\zeta\le-1$), but otherwise contributes to reduce the ensemble
 of anomalously-diffusive situations, as long as $\zeta>-1$. Such a reduction vanishes for $\zeta\to-1$
 (because there the constraint becomes again $\alpha>3-d$), and saturates to a maximum for $\zeta\ge1$ ($\Rightarrow\alpha>-1-d$).

\begin{figure}
 \centering
 \includegraphics{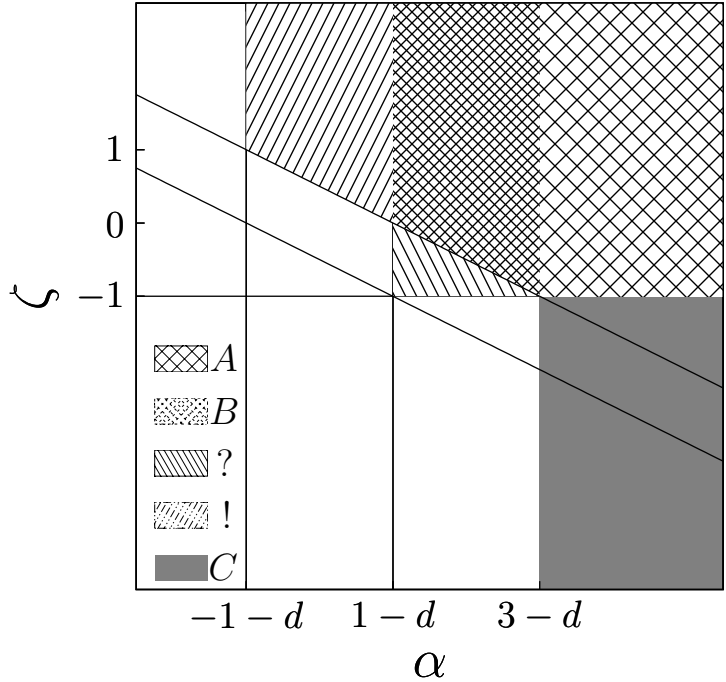}
 \caption{Sketch of diffusion anomaly for inertial particles, in the plane $\zeta$ vs.\ $\alpha$ (see (\ref{4}) coupled with (\ref{3})).
  Diffusion is definitely normal only in meshed areas. The coarse-mesh area A has normal diffusion
  also for steady flows, while in the fine-mesh area B the normality is induced by the time dependence.
  In lined areas either of the integrals (\ref{3}), (\ref{4}) diverges;
  namely, in zone ?\ only the inertial correction converges and anomaly is likely,
  while in zone !\ only the tracer contribution converges and an inertia-driven anomaly is possible.
  The union of A and of the grey area C indicates the definitely-normal zone for time-independent flows (see (\ref{0}) coupled with (\ref{1})).}
 \label{fig2}
\end{figure}

 The presence of inertia requires the study
\textsc{--- as an additive contribution ---} 
 of the integral in (\ref{corr}),
 $$\frac{\ell}{U}\!\int_0^Q\ud q\int_{-\Omega}^{\Omega}\ud\omega\,q^{d-2}q^{\alpha}|\omega|^{\zeta}\frac{2(1-\beta)\omega^2+(3-\beta)\kappa^2q^4}{2(\omega^2+\kappa^2q^4)}\;,$$
 which
\textsc{(for $\beta\neq1,3$)}
 establishes the following constraints for the scaling exponents:
 \begin{equation} \label{4}
  \alpha>1-d\quad\&\quad\zeta>-1\quad\&\quad\alpha+2\zeta>-1-d\;.
 \end{equation}
 The situation is sketched in figure \ref{fig2}.
 Some of these bounds, namely the last one, are overshadowed by stricter constraints from (\ref{3}).
 One can conclude that, neglecting higher orders in the Stokes-number expansion,
 for time-dependent flows situations do exist where tracers diffuse normally
 but inertial particle can diffuse anomalously: an example is provided by the case
 $\alpha=-d\ \&\ \zeta=1$
\footnote{\textsc{For the sake of simplicity, let us call it ``N-to-A situation''}}
 which satisfies (\ref{3}) but not (\ref{4}).
 (We say \emph{can diffuse}, and not ``do diffuse'', because we cannot exclude a regular renormalization for the full sum in (\ref{exp}).)
 In other words, the introduction of inertia may induce anomaly, but only for the cases $\zeta>0$,
 because then $1-d-2\zeta<1-d$ and the inertial bound on $\alpha$ in
\textsc{the first of}
 (\ref{4}) is more restrictive than the tracer counterpart from
\textsc{the last of}
 (\ref{3}).
\footnote{\textsc{From (\ref{corr}), it appears that different bounds hold
 for the specific cases $\beta=1$ and $\beta=3$.
 In the former case, the first bound in (\ref{4}) becomes $\alpha>-3-d$,
 and the aforementioned ``N-to-A situation'' is no longer possible.
 In the latter case, the second bound in (\ref{4}) becomes $\zeta>-3$,
 and the width of the normal region increases because temporal correlations are less crucial in causing anomaly.}}

 \subsubsection{Case of finite terminal velocity} \label{cvtf}

\begin{figure}
 \centering
 \includegraphics{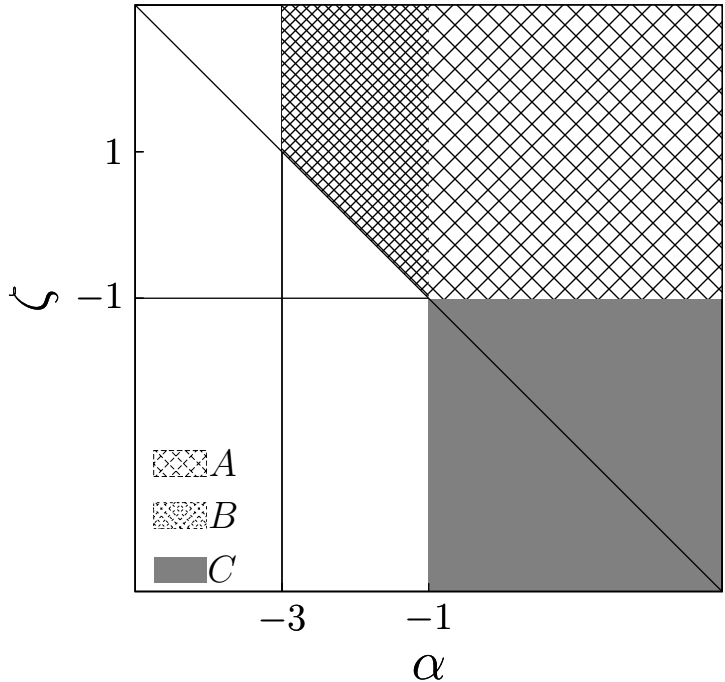}
 \caption{Sketch of diffusion anomaly for particles with finite terminal velocity, in the plane $\zeta$ vs.\ $\alpha$ (see (\ref{5})).
  Diffusion is normal only in meshed areas. The coarse-mesh area A has normal diffusion
  also for steady flows, while in the fine-mesh area B the normality is induced by the
  time dependence. The union of A and of the grey area C indicates the normal zone for time-independent flows (see (\ref{2})).}
 \label{fig3}
\end{figure}

\begin{figure}
 \centering
 \includegraphics{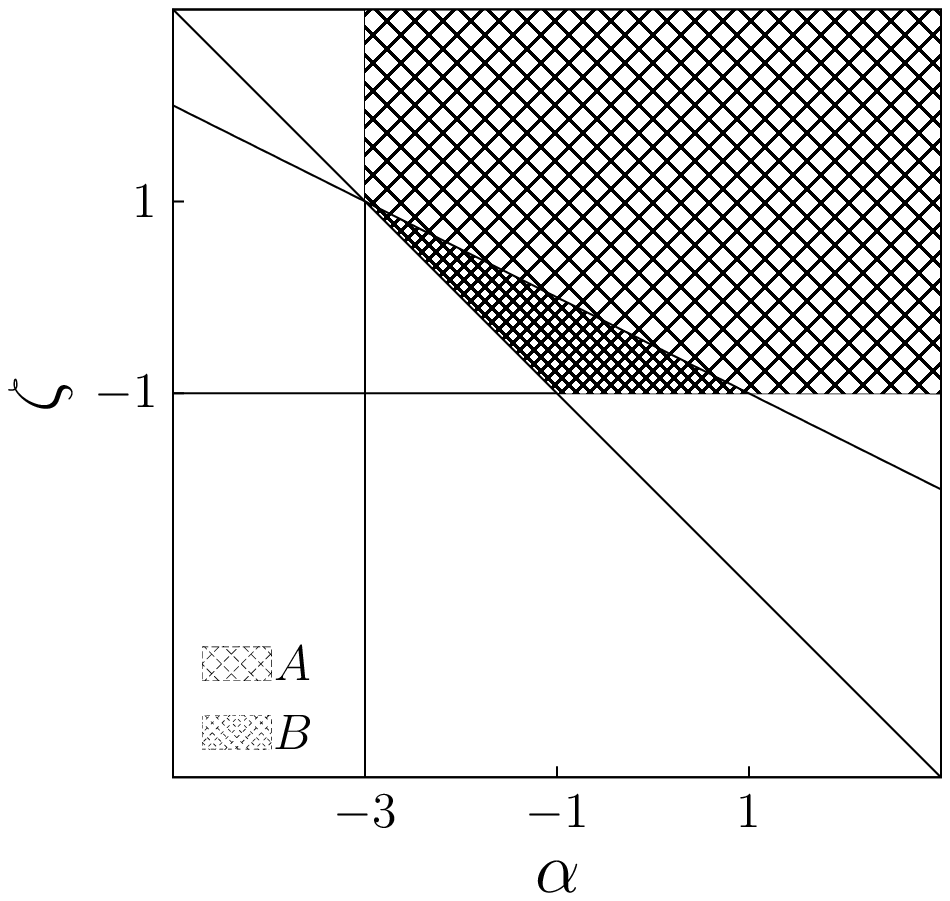}
 \caption{Sketch of diffusion anomaly for two-dimensional time-dependent flows, in the plane $\zeta$ vs.\ $\alpha$:
  comparison between the situations of vanishing and finite sedimentation (see (\ref{3}) and (\ref{5}), respectively,
  i.e.\ figures \ref{fig1} and \ref{fig3} for $d=2$).
  The coarse-mesh area A has normal diffusion in both cases, while in the fine-mesh area B normality only holds for finite sedimentation
  (no vice versa occurs).}
 \label{fig4}
\end{figure}

\begin{figure}
 \centering
 \includegraphics{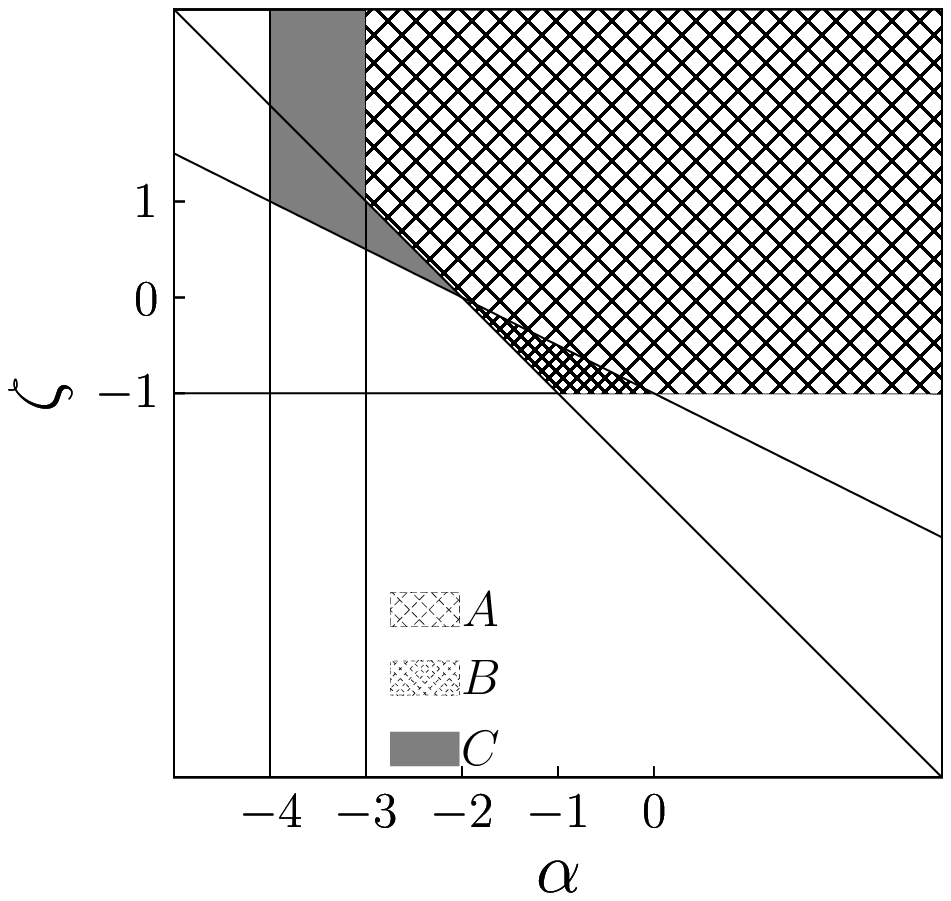}
 \caption{Sketch of diffusion anomaly for three-dimensional time-dependent flows, in the plane $\zeta$ vs.\ $\alpha$:
  comparison between the situations of vanishing and finite sedimentation (see (\ref{3}) and (\ref{5}), respectively,
  i.e.\ figures \ref{fig1} and \ref{fig3} for $d=3$).
  The coarse-mesh area A has normal diffusion in both cases, while in the fine-mesh area B normality only holds for finite sedimentation,
  and vice versa in the grey area C normality only holds for vanishing sedimentation.}
 \label{fig5}
\end{figure}

 As a last point, let us investigate the case of finite sedimentation for time-dependent flows.
 Even excluding the trivial cases of vertically-aligned flow and of neutrally-buoyant particles
 --- which would give back (\ref{trac}) and (\ref{3}) --- here the picture is complicated by the fact that
 the denominator of the integral in (\ref{max}) shows finite values of
 the parameters $q$ and $\omega$ for which the contribution
 in square brackets vanishes even for $d=2$, and the integrand thus changes its functional form (see discussion above).\\
 The result reduces to twice a double integral in the two-dimensional case:
 \begin{eqnarray*}
  \kappa\!\int_{-\Omega}^{\Omega}\ud\omega\left(\int_{-Q}^0\ud q\,q^{d-2}|q|^{\alpha}|\omega|^{\zeta}\frac{q^2}{[\omega+(1-\beta)\tau g\cos(\pi/2-\Theta)|q|]^2+\kappa^2q^4}\right.\\
  \left.+\!\int_0^Q\ud q\,q^{d-2}|q|^{\alpha}|\omega|^{\zeta}\frac{q^2}{[\omega+(1-\beta)\tau g\cos(\pi/2+\Theta)|q|]^2+\kappa^2q^4}\right)\\
  =2\kappa\!\int_{-\Omega}^{\Omega}\ud\omega\int_0^Q\ud q\,\frac{q^{\alpha+2}|\omega|^{\zeta}}{[\omega+(1-\beta)\tau g\sin(\Theta)q]^2+\kappa^2q^4}\;.\hspace{3.5cm}
 \end{eqnarray*}
 For $d=3$ the full triple integral is (with $\Re$ denoting the real part):
 \begin{eqnarray*}
  \kappa\!\int_{-\Omega}^{\Omega}\ud\omega\int_0^Q\ud q\int_0^{2\pi}\ud\theta\,q^{d-2}q^{\alpha}|\omega|^{\zeta}\frac{q^2}{[\omega+(1-\beta)\tau g\sin(\Theta)\cos(\theta)q]^2+\kappa^2q^4}\qquad\\
  =2\pi\!\int_{-\Omega}^{\Omega}\ud\omega\int_0^Q\ud q\,q^{\alpha+1}|\omega|^{\zeta}\Re\left\{\frac{1}{\sqrt{[(1-\beta)\tau g\sin(\Theta)q]^2-(\omega+\ui\kappa q^2)^2}}\right\}\;.
 \end{eqnarray*}
 In both cases, we obtain the following constraints:
 \begin{equation} \label{5}
  \alpha>-3\quad\&\quad\zeta>-1\quad\&\quad\alpha+\zeta>-2\;.
 \end{equation}
\textsc{As different comparisons are now possible, let us analyze separately the effects of time dependence and finite sedimentation.\\
I) Focusing on cases of particle finite sedimentation,
 let us investigate the role of the flow time dependence,
 by comparing (\ref{5}) with (\ref{2}).
 Both thresholds are independent of the dimension: see figure \ref{fig3}.
 (Notice that both thresholds would become dimension-dependent if expressed in terms of $\alpha'$ rather than $\alpha$.)
 If the temporal coherence is strong ($\zeta\le-1$) the diffusion is always anomalous.
 As long as $\zeta>-1$, the activation of a time dependence in the flow
 can transform anomalously-diffusive cases into standard ones:
 take e.g.\ $\alpha=-2$ and $\zeta=1$.
 The same remark had already been made when comparing (\ref{3}) with (\ref{1}),
 and thus holds for both vanishing and finite sedimentation.
 In other words, the introduction of a time dependence always induces anomaly
 in the presence of strong temporal coherence,
 but in the lack thereof it reduces the width of the anomalous region.\\
II) Focusing on cases of time-dependent flows,
 let us investigate the role of the particle sedimentation,
 by comparing (\ref{5}) with (\ref{3}).
 The comparisons for $d=2$ and $d=3$ are depicted in figures \ref{fig4} and \ref{fig5}, respectively.
 (Notice that all the thresholds not involving geometric arguments --- (\ref{1},\ref{0},\ref{3},\ref{4}) ---
 become independent of the spatial dimension if expressed in terms of the modified spectral exponent $\alpha'$.) 
 The activation of a particle finite sedimentation can both transform anomalously-diffusive situations into normal ones
 (take e.g.\ $\alpha=-1$ and $\zeta=-3/4$, for $d=2,3$), and vice versa (take e.g.\ $\alpha=-7/2$ and $\zeta=1$, \emph{only} for $d=3$).
 Only the former possibility had been remarked when comparing (\ref{2}) with (\ref{1}).
 This means that, not only for time-independent flows but also for \emph{two-dimensional} time-dependent ones,
 the introduction of a finite sedimentation \emph{reduces} the width of the anomalous region.
 However, for \emph{three-dimensional} time-dependent flows, a finite settling \emph{modifies}
 the shape of the anomalous region, so that areas of anomaly appear and other disappear.}

 \section{Conclusions and perspectives} \label{cope}
 
 In the context of the transport of weakly-inertial particles by a parallel flow,
 we have investigated their effective diffusivity to discriminate situations of normal or anomalous diffusion.
 In particular, we have studied the role of long-range spatial and temporal correlation
 in the fluid velocity spectrum, in terms of two parameters which determine the infrared behavior
 of the flow energy density. We have found different bounds identifying regions of anomaly or lack thereof,
 according to the properties of the particles (tracer or inertial), of the flow (steady or time-dependent)
 and of the suspension as a whole (vanishing or finite settling velocity).
 
\textsc{An interesting comparison arises between the present work and the formalism of fractional diffusion,
 where the equation under consideration is not our (\ref{de}) but rather
 \[\frac{\partial^{\eta}}{\partial T^{\eta}}P(\bm{X},T)=D\frac{\partial^{\Sigma}}{\partial X^{\Sigma}}P(\bm{X},T)\;,\]
 with real coefficients $\eta,\Sigma$, and suitable definitions of the Caputo and Riesz--Feller derivatives
 \cite{SW89,MGN94,YCST00,MLP01,ALEMM05,CSZK10,YJ13}. The underlying concept is that anomaly, in the sense of
 non-normality or non-Gaussianity, can be thought of as stemming from a ``grey'' noise,
 as opposed to the white noise corresponding to Brownian motion and ordinary diffusion.
 This point is not tackled here.}
 
 In this work we have only focused on the eddy diffusivity, i.e.\ the second-order moment of the particle dispersion.
 It would be interesting to also investigate higher-order moments, with the aim of understanding whether the anomalous diffusion
 is weak or strong. Namely, this latter adjective refers to the fact that not only the second moment is not asymptotically
 proportional to time, but also that higher moments exhibit different exponents which cannot be captured via a simple rescaling
 \cite{CMMV99,ACMV00}.
\textsc{For instance, in relation to particle dispersion in the terrestrial environment,
 no explicit parameterization for non-Gaussian behavior seems to be currently available
 in the state-of-the-art numerical modeling of this problem in the atmosphere.
 This paper might motivate new research toward this relevant direction.}

 One limitation of our work lies in the fact that its perturbative spirit makes it impossible to understand what happens
 when, in the small-inertia expansion truncated at the first order, either or both integrals diverge.
 In particular, in some occasions we could only assert that anomaly may arise due to this or that effect, but these simple hints
 should be verified or confuted by more in-depth analyses (using renormalization techniques) or by numerical simulations
 of the particle dynamics. Therefore, this paper represents a first step in the comprehension of anomalous diffusion
 when different physical effects are taken into account or neglected.
\textsc{In any case, we can assert that bounds (\ref{1},\ref{3}),
 as well as (\ref{2},\ref{5}) if settling is considered as independent of inertia,
 are exact results, and to our knowledge the last three are original in the scientific literature.}

 On the other hand, we are now developing a Lagrangian formalism to compute the effective diffusivity in parallel Kolmogorov flows
 without resorting to small-inertia expansions. In this way, at least for a specific class of flows,
 one can aim at generalizing the present results to inertial particles away from our perturbative limit of small Stokes number.
\textsc{Numerical simulations seem to represent the main tool for attacking non-parallel (for instance, cellular \cite{MAMMG12}) flows.}


 \appendix*
 \section{Regularization procedure and relation between exponents} \label{appe}

\textsc{A power-law infrared spectrum denotes the presence of excitations on scales arbitrarily large,
 which may put into question the validity of the multiple-scale formalism.
 A regularization procedure \cite{BCVV95} is then necessary, and is presented here for the case of steady flows.
 This consists in introducing an infrared cut-off length, $C_{\mathrm{spat}}$,
 and in replacing (in the integrals (\ref{trac},\ref{corr}--\ref{max})) the original spectrum $\mathcal{U}$ with
 \begin{equation} \label{reg}
  \bar{\mathcal{U}}(\vec{q},\omega)=\mathcal{U}(\vec{q},\omega)H(|\vec{q}|-C_{\mathrm{spat}}^{-1})\;,
 \end{equation}
 where $H$ denotes the Heaviside theta, killing the wave numbers smaller than the cut-off.
 The limit $C_{\mathrm{spat}}\to\infty$ is then taken only after performing the integrals,
 all of the type $\int_{C_{\mathrm{spat}}^{-1}}^Q$.\\
 This procedure has two important consequences.
 First, it may represent a way to reproduce this problem numerically,
 by studying the dependence of the (finite, for all finite $C_{\mathrm{spat}}$) eddy diffusivity
 on the cut-off length, and by performing the simulations in a box of such --- larger and larger --- size.\\
 Second, it is now possible to study the relation between the exponents $\gamma$ in (\ref{gam}) and $\alpha$ in (\ref{alfa}).
 The key point is the observation of the fact that the regime (\ref{gam}) is now expected at scales much larger than $\ell$ but
 sufficiently smaller than $C_{\mathrm{spat}}$ (potentially extending back to infinity in the aforementioned limit)
 \cite{MM80,YJ91,AM92,CV93}, while at spatial scales $\gg C_{\mathrm{spat}}$ --- and temporal scales
 \begin{equation} \label{tc}
  t\gg\bar{t}\equiv C_{\mathrm{spat}}^2/\kappa
 \end{equation}
 --- a truly-diffusive behavior holds:
 \begin{equation} \label{tdb}
  \langle|\bm{r}(t)-\bm{r}(0)|^2\rangle\sim K_{\lambda\lambda}(C_{\mathrm{spat}})t\;.
 \end{equation}
 By equating the mean square separations from (\ref{gam}) and (\ref{tdb}) at the cross-over time $\bar{t}$ in (\ref{tc}), one gets:
 \begin{equation} \label{match}
  (C_{\mathrm{spat}}^2)^{\gamma}\sim K_{11}(C_{\mathrm{spat}})C_{\mathrm{spat}}^2\;.
 \end{equation}
 The bounds (\ref{1}--\ref{2}) are now rephrased as $K^{(0)}_{11}\propto C_{\mathrm{spat}}^{-(\alpha+d-3)}$ for tracers
 (and $K^{(1)}_{11}\propto C_{\mathrm{spat}}^{-(\alpha+d-1)}$ for the leading inertial correction),
 and as $K^{(0)}_{11}\propto C_{\mathrm{spat}}^{-(\alpha+1)}$ for finitely-settling particles.
 Substituting these relations into (\ref{match}), at the leading order in the Stokes number one finally obtains
 \begin{equation} \label{ult}
  \gamma=\frac{5-d-\alpha}{2}\ \textrm{(vanishing settling)}\;,\qquad\gamma=\frac{1-\alpha}{2}\ \textrm{(finite settling)}\;,
 \end{equation}
 which are valid only when $\alpha\le3-d$ and $\alpha\le-1$ respectively, so that $\gamma\ge1$.}

\textsc{Moving to unsteady flows, in principle one should modify the regularization procedure for $\bar{\mathcal{U}}$ in (\ref{reg})
 by introducing an infrared cut-off $C_{\mathrm{temp}}$ also in the frequency domain,
 but the relation between $\gamma$ and $\alpha,\zeta$ is now more subtle than (\ref{ult}) and is not investigated here.}

 \bibliography{diffusanomalbis}

\end{document}